%% file: root.tex
\documentclass[conference,10pt]{IEEEtran}
\input{macros.tex}

\ifCLASSINFOpdf
\else
\fi
\begin{document}
\IEEEoverridecommandlockouts

\title{Enhanced Flexibility Aggregation Using LinDistFlow Model with Loss Compensation}



%
\author{\IEEEauthorblockN{Yanlin Jiang\IEEEauthorrefmark{2},
Xinliang Dai\IEEEauthorrefmark{2},
Frederik Zahn\IEEEauthorrefmark{2} and
Veit Hagenmeyer\IEEEauthorrefmark{2}}
\IEEEauthorblockA{\IEEEauthorrefmark{2}Institute for Automation and Applied Informatics\\
Karlsruhe Institute of Technology, Karlsruhe, Germany, 76131\\
Email: \{yanlin.jiang, xinliang.dai, frederik.zahn, veit.hagenmeyer\}@kit.edu}
\thanks{Corresponding: Xinliang Dai}
}



\maketitle
\begin{abstract}
With the increasing integration of renewable energy resources and the growing need for data privacy between system operators, flexibility aggregation methods have emerged as a promising solution to coordinate \acrfull{itd} systems with limited information exchange. However, existing methods face significant challenges due to the nonlinearity of AC power flow models, and therefore mostly rely on linearized models.
This paper examines the inherent errors in the LinDistFlow model, a linearized approximation, and demonstrates their impact on flexibility aggregation. To address these issues, we propose an intuitive compensation approach to refine the LinDistFlow-based flexibility set. Simulation results demonstrate the effectiveness of the proposed method in efficiently coordinating \acrshort{itd} systems.


\end{abstract}


%
\IEEEpeerreviewmaketitle

\section{Introduction}
In recent years, the share of \glspl{der} in global power generation has increased steadily. 
This transition poses significant challenges to traditional centralized coordination frameworks, primarily due to the growing complexity of system models introduced by the addition of \glspl{der}, raising privacy concerns and increased demands on communication~\cite{molzahn2017survey,patari2021distributed,cali2024emerging}.

To address these challenges, significant efforts have focused on developing solutions to efficiently coordinate modern \gls{itd} systems.
One promising approach is power flexibility aggregation, which leverages the collective flexibility of \glspl{der} to facilitate transmission-distribution interaction~\cite{dai2023nmpc,mathieu2024new,zhang2023coordination}. 

\begin{figure}
     \centering
     \includegraphics[width=0.85\linewidth]{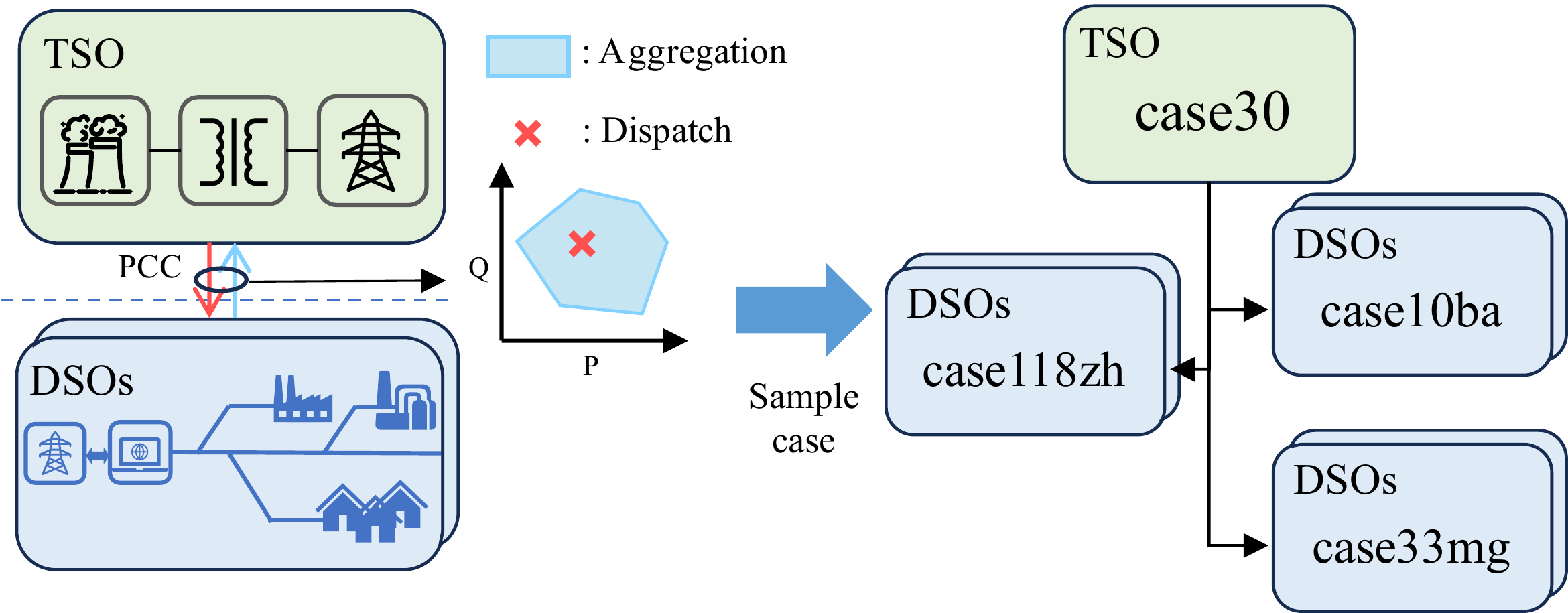}
     \caption{Coordination framework between TSO-DSO}
     \label{fig::coordination}\vspace{-9pt}
 \end{figure}

Flexibility refers to the set of all feasible power flows at the interface between \gls{tso} and \gls{dso}. By communicating this flexibility set, coordination problems, such as \gls{opf}, can be solved through a hierarchical coordination framework~\cite{wen2022aggregate}, as shown in \cref{fig::coordination}.

The primary challenge lies in the inherent nonlinearity of power system models. Existing flexibility aggregation methods often rely on linearized power flow models~\cite{tan2020estimating, zhang2023coordination}, with the LinDistFlow model commonly used~\cite{lopez2021quickflex,wen2023improvedDER}. By omitting line losses, the LinDistFlow model~\cite{baran1989network} can be viewed as a linearized version of the DistFlow model~\cite{baran1989optimal1,baran1989optimal2}, which serves as an exact angle relaxation of the AC model for radial distribution. However, these linearized models present significant issues: operational points within the flexibility set derived from linearized models may violate physical constraints in the AC model, posing risks to grid stability.

In the case of the AC model, where the flexibility boundaries are nonlinear, early aggregation methods relied on the Monte Carlo approach~\cite{riaz2019feasibility, contreras2019verification}. This method generates numerous operational scenarios and tests their feasibility. While straightforward to implement, this approach is computationally intensive and unreliable due to the randomness inherent in Monte Carlo sampling, which can result in poor-quality flexibility representations. 

As an alternative, direction-based sampling methods have been proposed~\cite{silva2018estimating, wang2024stochastic, kalantar-neyestanaki2020characterizing}. These methods optimize power at the \gls{pcc} to identify extreme boundary points along specific directions, offering a clearer visualization of the flexibility region. However, they do not streamline \gls{tso}-\gls{dso} coordination, as the boundary points are not expressed analytically. This requires additional verification steps to determine whether an operational point is within the feasible region. Furthermore, the computational complexity of these methods increases with the number of sampled points, adding to the overall computation time.

To balance the accuracy and computation burden, this paper focuses on improving the flexibility aggregation using the LinDistFlow model. The key contributions of this paper are as follows:
\begin{itemize}
\renewcommand\labelitemi{$\bullet$}
    \item Through an analysis of the LinDistFlow model, we demonstrate that the system errors omitted by the model accumulate at the coupling point, making it unsuitable for flexibility aggregation. This is illustrated through tutorial examples.  
    \item We propose a compensation method to address the system errors in the LinDistFlow model. The compensated power is represented as a quadratic mapping of power exchanges within the LinDistFlow-based flexibility set. This intuitive approach is both effective and computationally efficient, while preserving privacy by preventing the disclosure of detailed distribution grid information.
    \item Beyond tutorial examples, we extend the simulations to coordinated \acrfull{itd} systems, focusing on the impact of electricity price variations and scenarios involving multiple distribution systems. The results validate the proposed compensation method’s ability to improve accuracy with minimal additional computational effort.
\end{itemize}
The remainder of this article is structured as follows: Section~\ref{sec::model} introduces the power system model and the concept of coordination using the flexibility set. Section~\ref{sec::FA} presents the flexibility aggregation methods based on the LinDistFlow model and the proposed system loss compensation approach, illustrated with tutorial examples. Section~\ref{sec::sim} provides more extensive simulations to evaluate the proposed methods. Finally, Section~\ref{sec::conclusion} concludes the paper.

\section{Problem Statement}\label{sec::model}

Consider a power grid modeled as \acrfull{itd} system, represented by the tuple~$(\mathcal{N},\mathcal{L},\mathcal{S})$, where $\mathcal N$ represents the set of buses, $\mathcal{L}$ represents the set of branches, and $\mathcal S$ represents the set of subsystems, which include both transmission and distribution systems, i.e., $\mathcal{S} =  \mathcal{S}^\text{T}\cup \mathcal{S}^\text{D}.$

In this paper, we focus on a single transmission system, i.e., $\abs{\mathcal{S}^\text{T}}=1$, connected with multiple radial distribution systems, i.e., $\abs{\mathcal{S}^\text{D}}\geq1$. We assign the transmission system an index of $1$, i.e., $\mathcal{S} = \{1\}$. Each subsystem~$\xi\in\mathcal{S}$ is represented as a directed graph $(\mathcal{N}_\xi,\mathcal{L}_\xi)$, where $\mathcal{N}_\xi$ and $\mathcal{L}_\xi$ denote the set of buses and branches within the subsystem~$\xi$, respectively. For every radial distribution subsystem~$\xi\in\mathcal{S}^\text{D}$, the number of buses $N^\text{bus}_\xi$ and branches $N^\text{line}_\xi$ are related by $N^\text{bus}_\xi = N^\text{line}_\xi +1$. The power generation in the distribution systems refers to the controllable power output of distributed energy resources (DERs).

The transmission and each distribution system are interconnected at a \acrfull{pcc}, typically located at distribution substations corresponding to the first bus in the local bus set $\mathcal{N}_\xi$. In this paper, we assume that the voltage magnitude at the \acrshort{pcc} is fixed by the substations and is independent of the transmission system voltage. 

\subsection{Model of Transmission Systems}
In the transmission system ($\xi=1$), the complex voltage at a bus $i\in\mathcal{N}_1$ is expressed in polar coordinates, i.e., 
$V_i = v_i e^{\theta_i}$,
where $v_i$ and $\theta_i$ are the magnitude and angle of the complex voltage $V_i$. Thereby, the classic AC model of the transmission system can be written as follows
\begin{subequations}\label{eq::pf::transmission}
    \begin{align}
        p_i^g-p_i^\xi =v_i \sum_{k\in\mathcal{N}} v_k \left( g_{ik} \cos\theta_{ik} + b_{ik} \sin\theta_{ik} \right),&\forall i\in\mathcal{N}_1\label{eq::opf::pf::active}\\
            q_i^g-q_i^\xi =v_i \sum_{k\in\mathcal{N}} v_k \left( g_{ik} \sin\theta_{ik} -  b_{ik} \cos\theta_{ik} \right),&\forall i\in\mathcal{N}_1\label{eq::opf::pf::reactive}\\
        \underline v_i \leq v_i \leq \overline v_i,\;\;
        \underline p_i^g \leq p_i^g \leq \overline q_i^g,\;\;
        \underline q_i^g \leq q_i^g \leq \overline q_i^g,&\forall i\in\mathcal{N}_1\label{eq::opf::box}
\end{align}
\end{subequations}
where $p^g_i$ and $q^g_i$ (resp. $p^\xi$ and $q^\xi$) denote the real and reactive power produced by generators (resp. send to distribution system $\xi$) at bus $i$. 
These variables are set to 0 if no generator (resp. distribution system) is connected to a bus $i$.
The parameters $g$, $b$ denote the real and imaginary part of the complex nodal admittance matrix $Y$. The signs $\underline{\cdot}$ and $\overline{\cdot}$ denote upper and lower bounds for the corresponding state variables.

\subsection{Model of Distribution Systems}\label{sec::model::distribution}

In the following, we introduce two models for the distribution system: the DistFlow model~\cite{baran1989optimal1,baran1989optimal2} and the LinDistFlow model~\cite{baran1989network}. Both models are formulated using squared voltage coordinates, i.e., $U_i = \abs{V_i}^2,\;\forall i \in\mathcal{N}_{\xi}$.
\begin{rema}[\cite{farivar2013branch}]\label{rmk::DistFlow}
    The DistFlow model can be interpreted as an angle relaxation of the classical AC model~\eqref{eq::pf::transmission}, and the angle relaxation is exact when the distribution network is radial.
\end{rema}
\begin{rema} In the present paper, when we state that an approximation or relaxation is exact, we mean that the model produces results identical to those of the AC power flow model~\eqref{eq::pf::transmission}. \end{rema}

For clarity, we omit the subsystem index $\xi$ in the variables, as all variables belong to the same distribution system $\xi$. Furthermore, the system models are reformulated in matrix form for the methods introduced in the following section. Details on the connectivity matrices $C^t,\,C^f,\,\text{and}\,C^g$ can be found in~\cite{zimmerman2010matpower}, with the incidence matrix defined as $C = C^f-C^t$.
\subsubsection{DistFlow Model}  The DistFlow model can be written in a matrix form:
\begin{subequations}\label{eq::DistFlow}
    \begin{align}
        0 \,&=\,e_1^\top \, U - 1 \label{eq::DistFlow::ref} \\
        0 \,&=\, C \, U - 2(\mathbf{R} P_l + \mathbf{X} Q_l) + (\mathbf{R}^2+\mathbf{X}^2) \, L \label{eq::DistFlow::voltage}\\
        0 \,&=\, C^\top P_l + \parens{C^t}^\top \mathbf{R} \, L - e_1^\top p^\text{pcc} - C^g p^g + P^d\label{eq::DistFlow::active}\\
        0 \,&=\,C^\top Q_l + \parens{C^t}^\top \mathbf{X} \, L - e_1^\top q^\text{pcc} - C^g q^g + Q^d\label{eq::DistFlow::reactive}\\
        0 \,&=\, \mathbf{U}\, L -  \left( \mathbf{P}^2 +  \mathbf{Q}^2\right) \;  \label{eq::DistFlow::quadratic}\\
        \underline{U} \,&\leq \, U\leq \overline{U},\;\underline{p}^\text{pcc} \leq  p^\text{pcc}\leq \overline{p}^\text{pcc},\;\underline{q}^\text{pcc} \leq  q^\text{pcc}\leq \overline{q}^\text{pcc}\label{eq::DistFlow::bound2}\\\
        \underline{p}^g &\leq  p^g\leq \overline{p}^g,\;\underline{q}^g \leq  q^g\leq \overline{q}^g\label{eq::DistFlow::bound1}
    \end{align}
\end{subequations}
with
\begin{align*}
    &\mathbf{P} = \textrm{diag}(P_l),\; &\mathbf{Q} = \textrm{diag}(Q_l),\;&\mathbf{U} = \textrm{diag}(C\,U),\\&\mathbf{R}=\textrm{diag}(R),\;&\mathbf{X}=\textrm{diag}(X),\;&
\end{align*}
where $e_1 = [1,0,\cdots,0]^\top\in\mathbb{R}^{N^\textrm{bus}}$, and $U\in\mathbb{R}^{N^\textrm{bus}}$ stacks squared voltage magnitudes for all buses in $\mathcal{N}$. The vectors $L,\,P_l,\,Q_l,\,X,\,R\in\mathbb{R}^{N^\textrm{line}}$ represents stacked squared current magnitude $l_{ij}$ and power flows $p_{ij}$, $q_{ij}$, resistance $r_{ij}$ and reactance $x_{ij}$ of branch $(i,j)\in\mathcal{L}_\xi$, respectively. The vectors $P^g$ and $Q^g$ (resp.  $P^d$ and $Q^d$) denote vectors stacked with the active and reactive generation injections of controllable devices (resp. demands, i.e., consumption by customer). $p^\textrm{pcc}$ and $q^\textrm{pcc}$ denote active and reactive power injections via \acrshort{pcc} from a transmission system.

Additionally, due to power conservation in the subsystem~$\xi$, we have: 
\begin{subequations}\label{eq::conservation}
    \begin{align}
        p^\text{pcc}+ \sum_{i\in\mathcal{N}_\xi} p^g_i - \sum_{i\in\mathcal{N}_\xi} p^d_i - R^\top L &=0,\\
        q^\text{pcc}+ \sum_{i\in\mathcal{N}_\xi} q^g_i - \sum_{i\in\mathcal{N}_\xi} q^d_i - X^\top L &=0.
    \end{align}
\end{subequations}


\subsubsection{LinDistFlow Model} The LinDistFlow model can be expressed in matrix form as:
\begin{subequations}\label{eq::LinDistFlow}
    \begin{align}
        0 \,&=\,e_1^\top \, U - 1 \\
        0 \,&=\, C \, U - 2(\mathbf{R} P_l + \mathbf{X} Q_l)  \\
        0 \,&=\, C^\top P_l  - e_1 p^\text{pcc} - C^g p^g + P^d\\
        0 \,&=\,C^\top Q_l  - e_1 q^\text{pcc} - C^g q^g + Q^d\\
        \underline{U} \,&\leq \, U\leq \overline{U},\;\underline{p}^\text{pcc} \leq  p^\text{pcc}\leq \overline{p}^\text{pcc},\;\underline{q}^\text{pcc} \leq  q^\text{pcc}\leq \overline{q}^\text{pcc}\\\
        \underline{p}^g &\leq  p^g\leq \overline{p}^g,\;\underline{p}^g \leq  q^g\leq \overline{p}^g
    \end{align}
\end{subequations}
\begin{rema}[\textbf{Lossless Approximation}]\label{rmk::lossless}
    The LinDistFlow~\eqref{eq::LinDistFlow} is derived by omitting higher-order power loss terms in \eqref{eq::DistFlow::voltage}\eqref{eq::DistFlow::active}\eqref{eq::DistFlow::reactive}, i.e., setting $l_{ij}=0$ for all branches $(i,j)\in\mathcal{L}_\xi$. This omission also removes the quadratic constraint~\eqref{eq::DistFlow::quadratic}, making LinDistFlow a linearized version of the DistFlow model~\eqref{eq::DistFlow} that ignores power losses. 
\end{rema}

\subsection{Optimal Power Flow for ITD Systems}
To ensure consistency, the power injected into the system at the \acrshort{pcc} must equal the power supplied from the transmission system at the corresponding bus $i$, as expressed by:
\begin{equation}\label{eq::consensus}
    p^\text{pcc}_\xi= p_i^\xi ,\; q^\text{pcc}_\xi= q_i^\xi,
\end{equation}
where $i\in\mathcal{N}_1$. 
The objective is to minimize the overall power generation cost defined as
\begin{equation}\label{eq::of::cost}
    \sum_{i\in \mathcal{N}}{c_{i,2}\;(p_{i}^g)^2+c_{i,1}\; p_{i}^g+ c_{i,0}}
\end{equation}
where $a_{i,2}$, $a_{i,1}$, and $a_{i,0}$ denote the cost coefficients of operation cost of power generations at bus $i$. 

The \acrfull{opf} problems for the \acrshort{itd} systems can be written as
\begin{subequations}\label{eq::opf::itd}
    \begin{flalign}
        \qquad \qquad \qquad \min \; &\eqref{eq::of::cost}&\text{(Generation Cost)}\\
        \text{s.t.}\;&\eqref{eq::pf::transmission},&\text{(Trans. Constraints)}\\
        & \eqref{eq::DistFlow},\;\forall \xi\in\mathcal{S}^\text{D},\;&\text{(Distr. Constraints)}\label{eq::opf::itd::distribution}\\
       & \eqref{eq::consensus},\;\forall\xi\in\mathcal{S}^\text{D}&\text{(Power Exchange)}\label{eq::opf::itd::exchange}
    \end{flalign}
\end{subequations}
The solutions to~\eqref{eq::opf::itd} are exact when all distribution networks are radial, as noted in Remark~\ref{rmk::DistFlow}.

Flexibility aggregation methods are employed to define a feasible region $\mathcal{U}^\xi$ for each distribution system~$\xi\in\mathcal{S}^\text{D}$. This allows the optimization problem \eqref{eq::opf::itd} to be reformulated as:
\begin{subequations}\label{eq::opf::itd::flex}
    \begin{flalign}
        \qquad  \min \quad &\eqref{eq::of::cost}\;     \\  \text{s.t.}\quad&\eqref{eq::pf::transmission},\\
        & (p^\text{pcc}_\xi,q^\text{pcc}_\xi)\in\mathcal{U}^\xi,\;\forall\xi\in\mathcal{S}^\text{D},
    \end{flalign}
\end{subequations}
so that detailed constraints for distribution systems \eqref{eq::opf::itd::distribution} are no longer required.

\section{Flexibility Aggregation Methods}\label{sec::FA}
In this section, we begin by discussing flexibility aggregation methods based on the LinDistFlow model~\eqref{eq::LinDistFlow}. We then illustrate the challenges associated with these methods and introduce an intuitive compensation approach based on the properties of the model with tutorial examples. Similar to Section~\ref{sec::model::distribution}, we omit the subsystem index $\xi$ in this section.

\subsection{Flexibility Aggregation based on LinDistFlow Model}\label{sec::FA::LinDistFlow}
The LinDistFlow models~\eqref{eq::LinDistFlow} can be reformulated as follows:
\begin{equation}
\label{eq::lindistflow:short}
    A x + B u - b = 0,\;\underline{x}\le x\le\overline{x},\;\underline{u}\le u\le\overline{u}
\end{equation}
with
\[
    A \,=\,\begin{bmatrix}
        e_1^\top & 0 & 0 & 0 & 0\\
        C  & -2\mathbf{R} & -2\mathbf{X} & 0 &0 \\
        0 &  -C^\top & 0 & C^g & 0 &   \\
        0 & 0 & -C^\top & 0 & C^g
    \end{bmatrix},\quad B= \begin{bmatrix}
        0 & 0\\
        0 & 0\\
        e_1 & 0\\
        0 & e_1
    \end{bmatrix},\\
\]
$x^\top\,=\,\begin{bmatrix}
        U^\top&\parens{P^l}^\top&\ \parens{Q^l}^\top & p^g &q^g
    \end{bmatrix},\\u^\top=\begin{bmatrix}
        p_{\mathrm{pcc}}&q_{\mathrm{pcc}}
    \end{bmatrix}^\top\;\text{and}\;\; b = \begin{bmatrix}
        e_1^\top&0& \parens{P^d}^\top& \parens{Q^d}^\top 
    \end{bmatrix}.$
\begin{propo}[\textbf{Prop.~1 in \cite{dai2023nmpc}}]\label{prop::invertible}
Given a radial network, matrix $A$ is invertible.
\end{propo}
\begin{rema}
This paper focuses on improving flexibility aggregation methods to address the challenges of nonlinearity in the AC model. Benchmarks used in this paper are limited to a single \acrshort{der} with controllable active and reactive power injections in one distribution system. 
For systems with multiple controllable \acrshort{der}, participation factors, as used in system balancing~\cite{lee2021robust}, are recommended to ensure invertibility to the matrix $A$.
\end{rema}
The task in flexibility aggregation is to propagate the constraints on $x$, through the grid and map them onto the active and reactive power exchange $u$ at the \acrshort{pcc}. For the LinDistFlow model~\eqref{eq::lindistflow:short}, since the matrix $A$ is invertible, all state variables $x$ can be explicitly expressed in terms of $u$ as follows:
\begin{equation}\label{eq::LinDistFlow::x}
    x =  A^{-1}\parens{b-Bu},
\end{equation}
and the resulting flexibility set can be expressed as
$$\mathcal{U}_\textsc{lds} = \left\{u\,|\, \underline{x}\le A^{-1}\parens{b-Bu}\le\overline{x},\;\underline{u}\le u\le\overline{u} \right\}.$$
However, omitting power losses (Remark~\ref{rmk::lossless}) introduces errors in the system. Compared to other applications, such as local voltage control strategies~\cite{farivar2013equilibrium}, aggregating the flexibility in a distribution system requires defining a broader set at the \acrshort{pcc}. Moreover, all line losses are effectively accumulated at $(p^\text{pcc},q^\text{pcc})$, causing the errors introduced by the LinDistFlow model to propagate into the resulting flexibility set $\mathcal{U}_\textsc{lds}$.

To illustrate this, three tutorial examples, i.e., case10ba, case33mg, and case118zh from \textsc{Matpower}~\cite{zimmerman2011matpower}, are shown in Fig.~\ref{fig::aggr::3cases}. The topologies of these examples are presented in the first row of Fig.~\ref{fig::aggr::3cases}, demonstrating that all \acrshort{der}s are connected to one of the leaf nodes, thereby accumulating all line losses along the path to the root node.

\Cref{fig::aggr::118zh}, \Cref{fig::aggr::10ba}, and \Cref{fig::aggr::33mg} depict the exact flexibility sets bounded by voltage limits (blue lines), active power limits (red lines), and reactive power limits (purple lines). Solid lines represent the upper bounds, while dashed lines correspond to the lower bounds. The red region represents the LinDistFlow-based flexibility set $\mathcal{U}_\textsc{lds}$. Notably, the flexibility set $\mathcal{U}_\textsc{lds}$ exhibits relatively large errors due to the accumulation of losses, and these errors grow as the system size increases.

\subsection{System Loss Compensation}\label{sec::FA::compensation}

To address the errors introduced by the LinDistFlow model, we propose an intuitive approach to compensate for system losses inspired by power conservation~\eqref{eq::conservation}. To simplify computation, we first introduce the following assumption:
\begin{ass}\label{ass::voltage}
     We assume that the voltage difference across branches remains constant within the flexibility set $\mathcal{U}$.
\end{ass}
The main idea is to first determine the LinDistFlow-based flexibility set~$\mathcal{U}_\textsc{lds}$, as described in Section~\ref{sec::FA::LinDistFlow}. Since all state variables $x$ can be expressed as a linear mapping from $u$ via~\eqref{eq::LinDistFlow::x}, we have: 
\begin{subequations}
\begin{align}
    \mathbf{P}(u) &= \text{diag}\parens{\bar{A}_p\;u + \bar{b}_p}\\
    \mathbf{Q}(u) &= \text{diag}\parens{\bar{A}_q\;u + \bar{b}_q}
\end{align}
\end{subequations}
where $\bar{A}_p,\bar{A}_q$ and $\bar{b}_p, \bar{b}_q$  are constructed from the corresponding rows of $-A^{-1}B$ and $A^{-1}b$ in~\eqref{eq::LinDistFlow::x}.

\begin{figure*}[htbp]
    \centering
    \begin{subfigure}[t]{0.3\textwidth}
        \caption{Topology of case10ba}
        \vspace{3pt}
        \includegraphics[width=\textwidth]{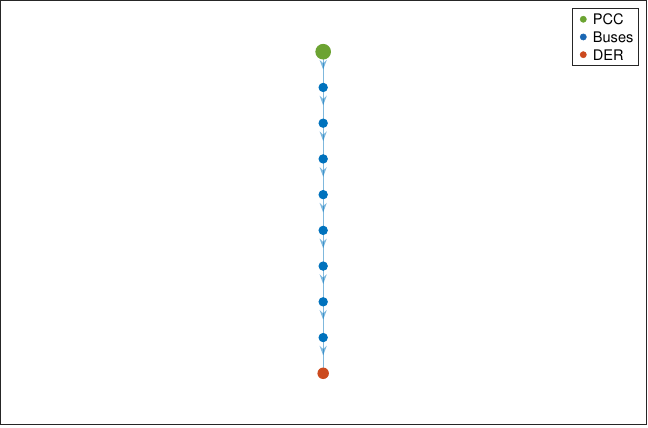} 
        \label{fig::topo::10ba}
    \end{subfigure}
    \hfill
    \begin{subfigure}[t]{0.3\textwidth}
        \caption{Topology of case33mg}
        \vspace{3pt}
        \includegraphics[width=\textwidth]{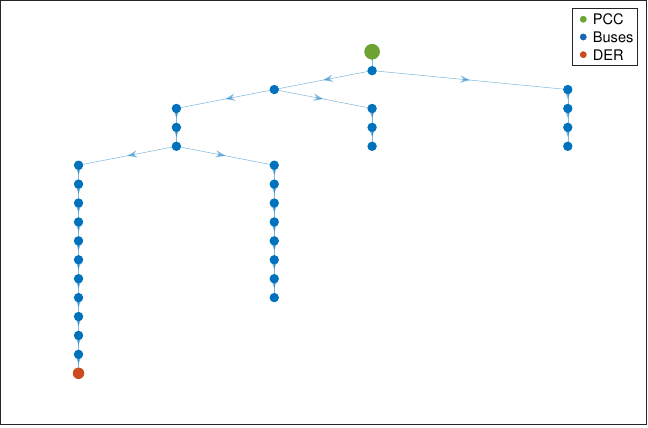} 
        \label{fig::topo::33mg}
    \end{subfigure}
    \hfill
    \begin{subfigure}[t]{0.3\textwidth}
        \caption{Topology of case118zh}
        \vspace{3pt}
        \includegraphics[width=\textwidth]{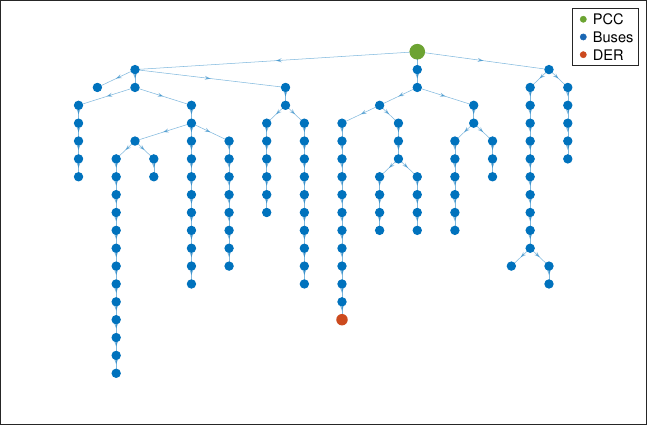} 
        \label{fig::topo::118zh}
    \end{subfigure}

    \vspace{-3pt}
    
    \begin{subfigure}[t]{0.3\textwidth}
        \centering
        \caption{Aggregation for case10ba}
        \includegraphics[width=\textwidth]{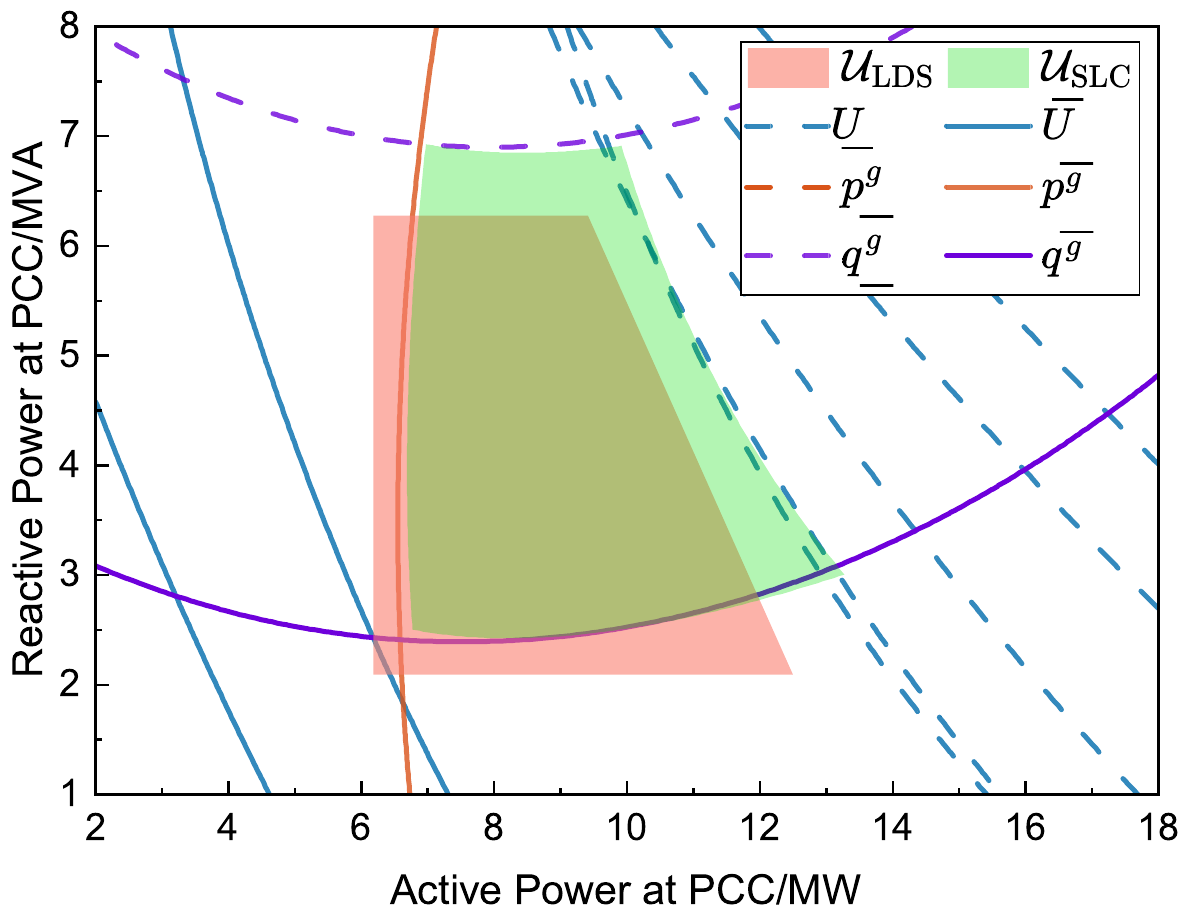} 
        \label{fig::aggr::10ba}
    \end{subfigure}
    \hfill
    \begin{subfigure}[t]{0.305\textwidth}
        \centering
        \caption{Aggregation for case33mg}
        \includegraphics[width=\textwidth]{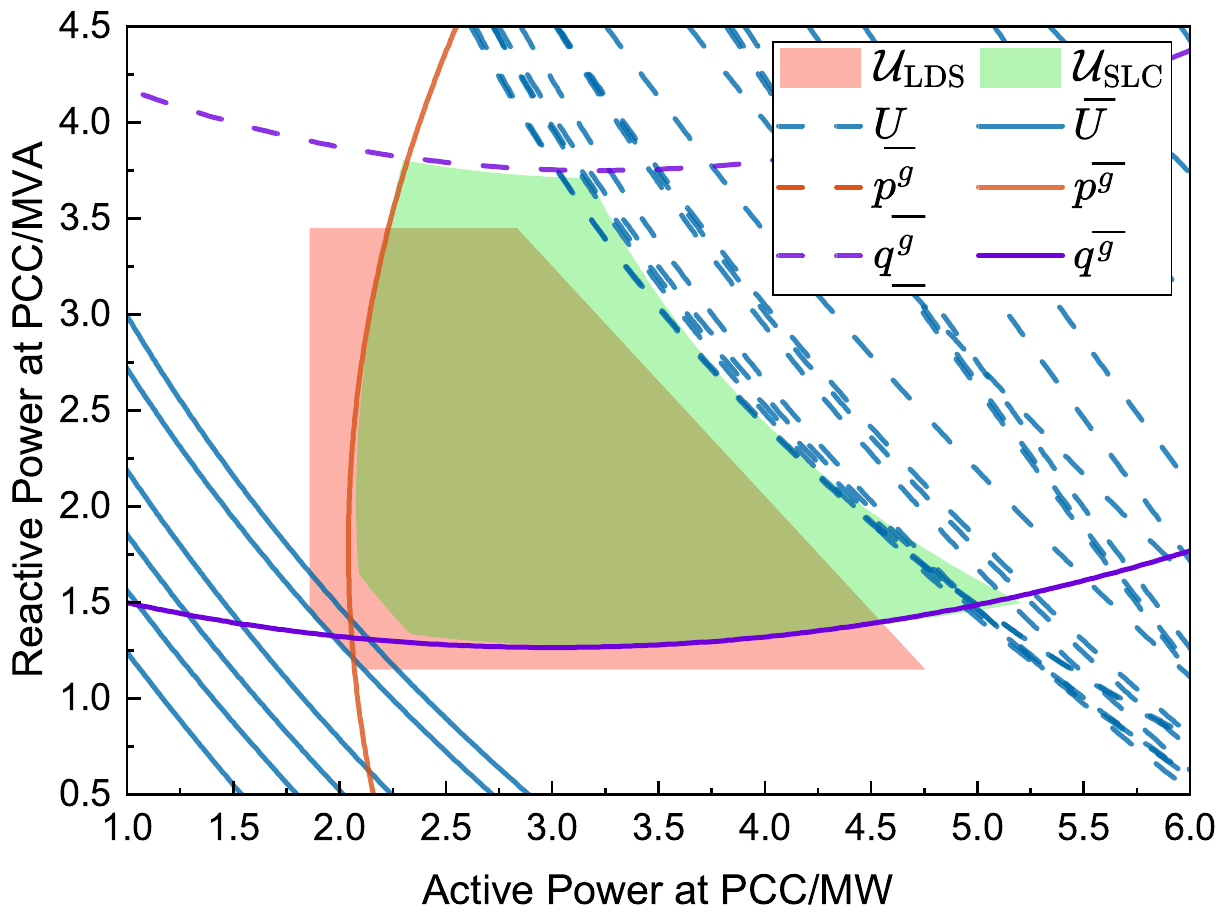}
        \label{fig::aggr::33mg}
    \end{subfigure}
    \hfill
    \begin{subfigure}[t]{0.3\textwidth}
        \centering
        \caption{Aggregation for case118zh}
        \includegraphics[width=\textwidth]
        {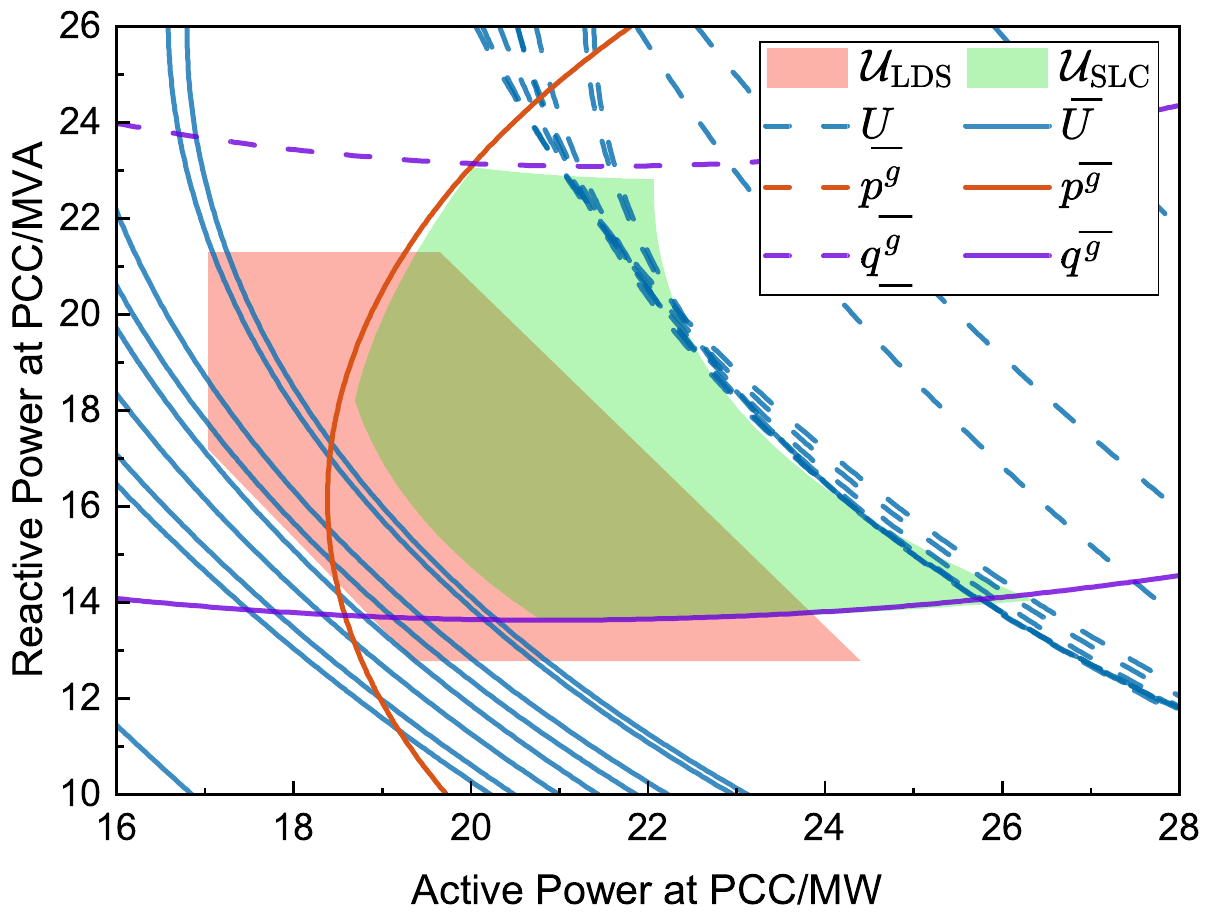} 
        \label{fig::aggr::118zh}
    \end{subfigure}

    \vspace{-9pt}
    
    \begin{subfigure}[t]{0.3\textwidth}
        \centering
        \caption{Dispatch tests for 10ba}
        \includegraphics[width=\textwidth]{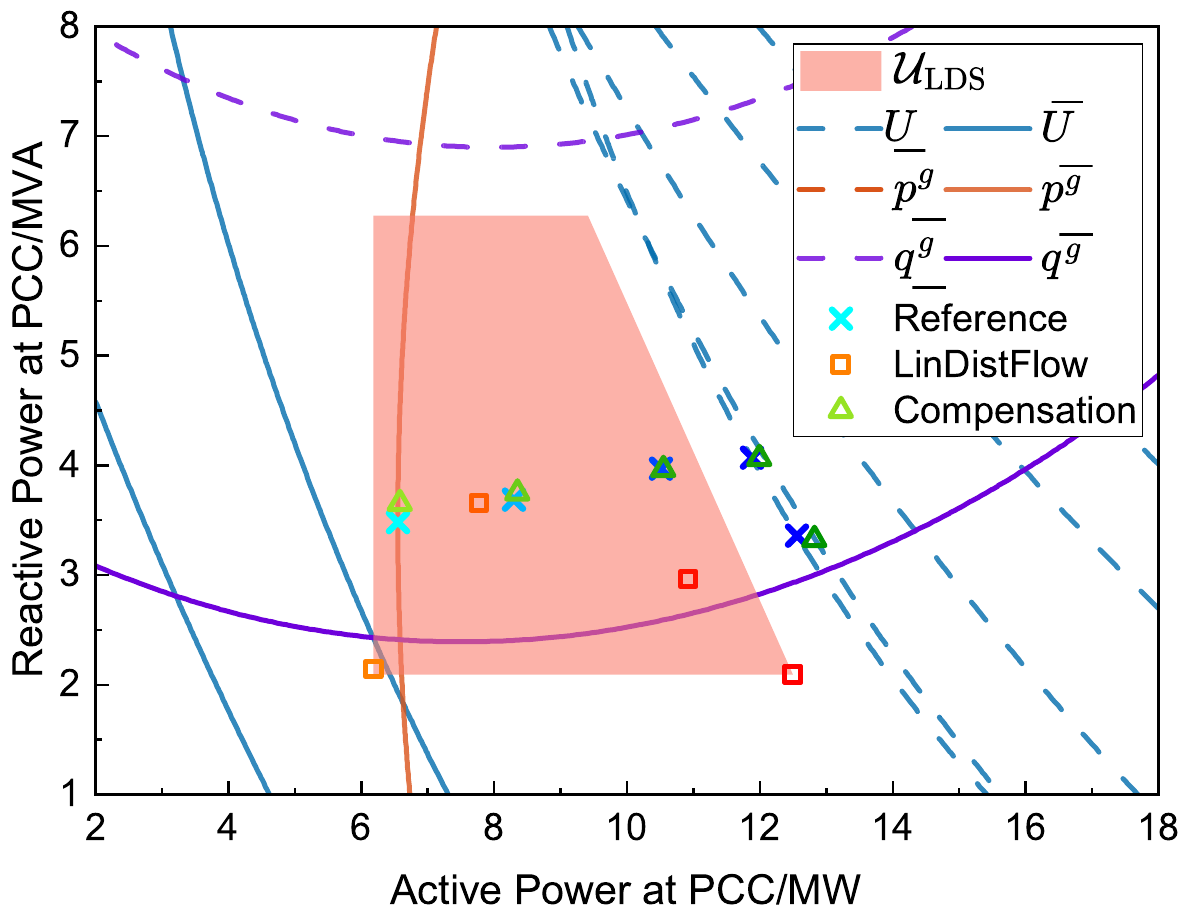} 
        \label{fig::dispatch::10ba}
    \end{subfigure}
    \hfill
    \begin{subfigure}[t]{0.305  \textwidth}
        \centering
        \caption{Dispatch tests for 33mg}
        \includegraphics[width=\textwidth]{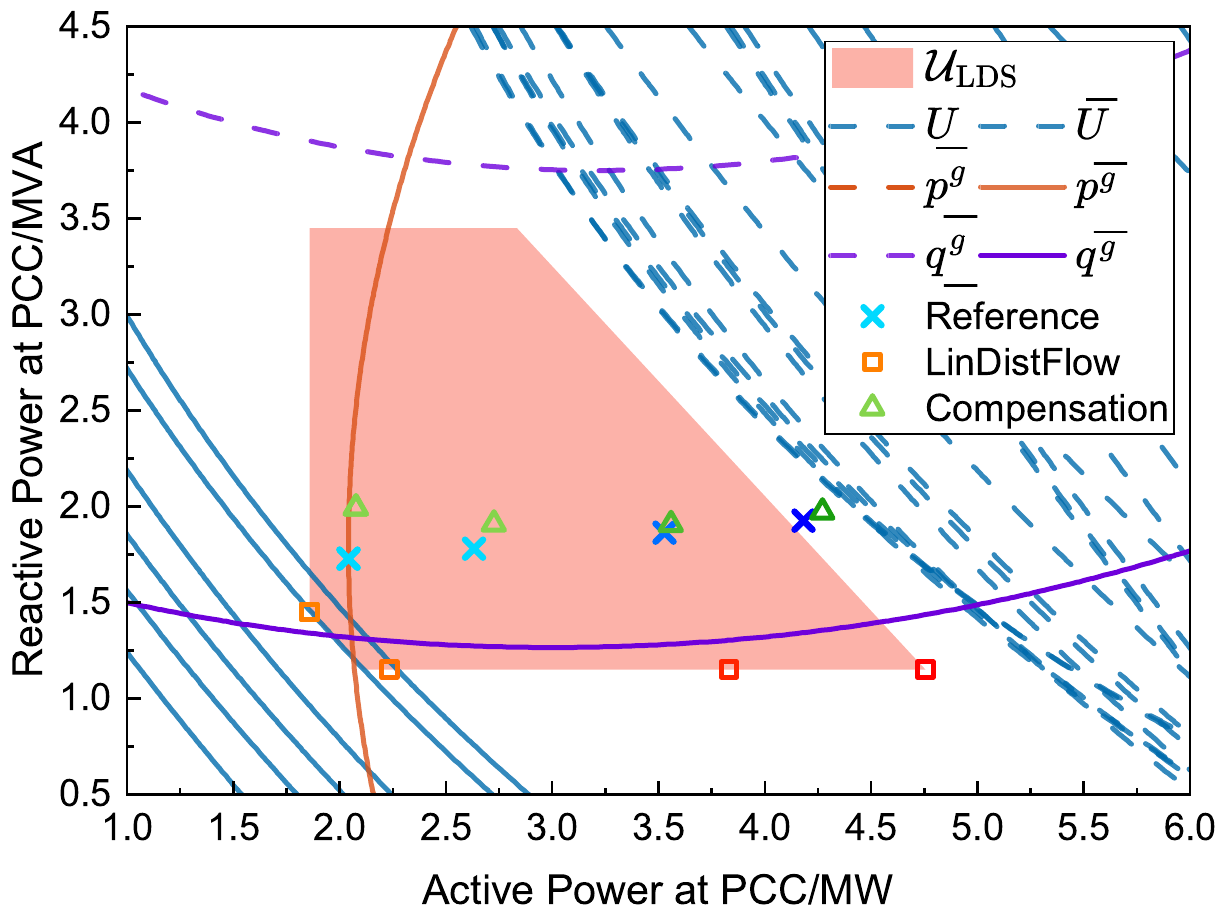} 
        \label{fig::dispatch::33mg}
    \end{subfigure}
    \hfill
    \begin{subfigure}[t]{0.3\textwidth}
        \centering
        \caption{Dispatch tests for case118zh}
        \includegraphics[width=\textwidth]{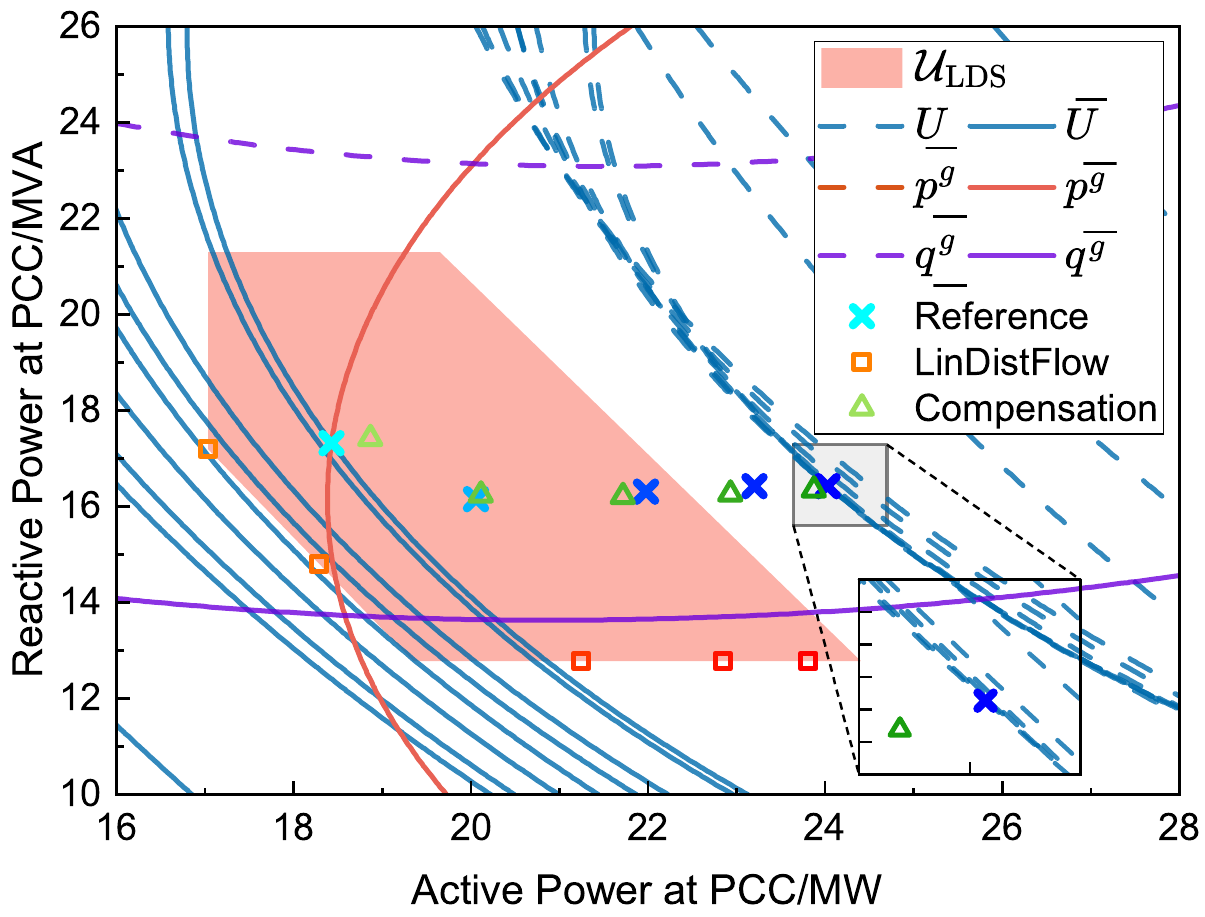} 
        \label{fig::dispatch::118zh}
    \end{subfigure}    
    \vspace{-9pt}
    \caption{First row: The topology of the radial networks, with DER location and PCC. Second row: Shaded regions represent the LinDistFlow model (red) (i) and the compensation method (green) (ii). Dashed and solid lines mark the exact flexibility bounds, derived from solutions using DistFlow model. Third row: Markers show \gls{pcc} power flows for different costs, increasing in price from left to right.}
    \label{fig::aggr::3cases}\vspace{-12pt}
\end{figure*}

With Assumption~\ref{ass::voltage}, system losses can be efficiently computed as quadratic functions of $u$:
\begin{subequations}\label{eq::loss}
    \begin{align}
        p^\text{loss}(u) &= R^\top \hat{\mathbf{U}}^{-1} \parens{\mathbf{P}^2(u)+\mathbf{Q}^2(u)}\,\mathds{1},\\
        q^\text{loss}(u) &= X^\top \hat{\mathbf{U}}^{-1} \parens{\mathbf{P}^2(u)+\mathbf{Q}^2(u)}\,\mathds{1},
    \end{align}
\end{subequations}
where $\mathds{1}\in\mathbb{R}^{N^\text{line}}$ denotes the vector of all ones, and $\hat{\mathbf{U}} = \text{diag}(C\,\hat{U}) = \text{diag}(C^f\,\hat{U}-C^t\,\hat{U})$ represents an estimate of the squared voltage differences. In this paper, we compute $\hat{\mathbf{U}}$ by setting $p^g= q^g=0.$
Alternatively, the losses can be reformulated as:
\begin{subequations}\label{eq::loss::quadratic}
    \begin{align}
        p^\text{loss}(u) &= \frac{1}{2} u^\top H_p u+ g_p^\top u + c_p,\\
        q^\text{loss}(u) &= \frac{1}{2} u^\top H_q u+ g_q^\top u + c_q.
    \end{align}
\end{subequations}
with $H_p,\,H_q\in\mathbb{R}^{2\times2}$, $g_p,g_q\in\mathbb{R}^2$ and $c_p,c_q\in\mathbb{R}$.
\begin{rema}[\textbf{Data Privacy}]
    The quadratic functions $p^\text{loss}(\cdot), q^\text{loss}(\cdot)$ accumulate estimated losses across all branches. As a result, transmitting $H_p,\,H_q,\,g_p,g_q,\,c_p,\,c_q$ does not reveal sensitive data.
\end{rema}
The flexibility set with system loss compensation can then be written as:
\begin{equation*}
    \mathcal{U}_\text{SLC} = \left\{ u\,\middle\vert\,  u = u_\text{LDS}+\begin{bmatrix}
        p^\text{loss}(u_\text{LDS})\\
        q^\text{loss}(u_\text{LDS})
    \end{bmatrix},\;u_\text{LDS} \in\mathcal{U}_\text{LDS} \right\}.
\end{equation*}
The effectiveness of the proposed compensation is illustrated in Fig.~\ref{fig::aggr::3cases}. The flexibility sets with system loss compensation $\mathcal{U}_\text{SLC}$ are shown as the green regions in the subfigures of the second row. Compared to the original LinDistFlow-based flexibility set $\mathcal{U}_\text{LDS}$, the compensation significantly improves the approximation of the exact flexibility set. Although some violations exist—such as the upper-right corner of for example, the upper-right corner of $\mathcal{U}_\text{SLC}$ in Fig.~\ref{fig::aggr::118zh} and relatively large deviations near lower voltage limits—these may result from the constant voltage difference assumption (Assumption~\ref{ass::voltage}) used for computational efficiency. Nevertheless, the compensation provides a good approximation for power limits.

\section{Simulation}\label{sec::sim}
We evaluate our loss compensation approach through numerical simulations of a coordination problem. Specifically, we assess the solution quality by solving the flexibility optimization problem~\eqref{eq::opf::itd} for distribution systems using both the LinDistFlow-based flexibility aggregation and the proposed loss compensation method. The results are compared to those of the original optimization problem~\eqref{eq::opf::itd::flex}.

The simulations include two scenarios: the first involves a single \gls{dso} with varying generation costs, while the second addresses a coordination problem with 29 distribution systems. We use radial distribution networks provided by \textsc{Matpower} as the test systems, each equipped with a \acrfull{der}. A controllable \gls{der} is placed at a leaf node in each network, as shown in Fig.~\ref{fig::aggr::3cases}. The power bounds of the \glspl{der} are set as fixed percentages of the total demand.


\subsection{Performance with Different Cost}
To evaluate coordination with aggregated flexibility~\eqref{eq::opf::itd::flex} under various scenarios, we first consider the \gls{tso} connected to a single \gls{dso}. We simulate three different cases by choosing three distribution grids, namely case118zh, case10ba, and case33mg. In electricity markets, prices fluctuate over time, and we simulate different coordination scenarios by varying the cost parameters of the \gls{dso}, ranging from lower to higher than those of the \gls{tso}. For each scenario, the corresponding power injection at the \gls{pcc} is recorded.

As a reference for comparison, dispatch points are first calculated solving the centralized coordination problem~\eqref{eq::opf::itd}, where the \gls{tso} is modeled with the AC power flow model, and the \gls{dso} is modeled with the DistFlow model. These results, marked as blue crosses in \Cref{fig::dispatch::118zh}-\Cref{fig::dispatch::33mg}, are arranged from left to right, corresponding to increasing prices.

Next, we solve the coordination problem with aggregated flexibility~\eqref{eq::opf::itd::flex}. When using the LinDistFlow model, the results are marked as red rectangles. Because the LinDistFlow model neglects power losses, these dispatch points are positioned at the boundaries of the exact flexibility region, violating the limits established by the DistFlow model. However, applying the compensation method corrects for these power losses by adding them to the power flow at the \gls{pcc}. The resulting dispatch points, marked as green triangles, align closely with the reference points, demonstrating the effectiveness of the compensation method.

\subsection{Overall Performance with Multiple Distribution Systems}
In the scenarios discussed above, the compensation method demonstrates that decentralized coordination can closely approximate the exact dispatch decisions. Now, we consider a more complex scenario where multiple networks are interconnected. This merged system consists of a transmission network (case30) and three types of distribution networks—case118zh, case10ba, and case33mg—operated by \glspl{dso}, following the configurations described above. The connectivity matrix $C_\mathrm{pcc}$ is constructed by sorting the active demands in case30: nodes with demands exceeding 15 MW are connected to case118zh, those between 5 MW and 15 MW to case10ba, and those below 5 MW to case33mg, as illustrated in \cref{fig::coordination}.

Decentralized coordination determines only the power flow at the \gls{tso}-\gls{dso} interface, leaving the internal states of the distribution networks unresolved. To address this, each \gls{dso} must perform a post-coordination calculation, solving a power flow problem using the fixed power at the \gls{pcc} as input. It ensures that \glspl{der} within the networks are scheduled appropriately and provides operational data for the \glspl{dso}.

\begin{figure}[htbp!]
\centering
\begin{subfigure}[t]{0.4\textwidth}
    \centering
    \caption{Flexibility aggregation using the LinDistFlow model}\label{fig::performance::LDF} 
    \includegraphics[width=\textwidth]{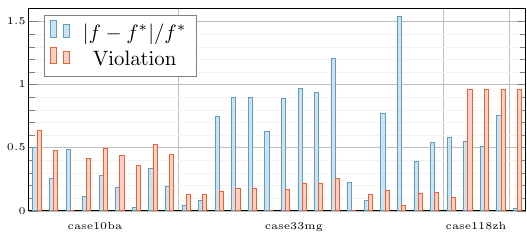}
\end{subfigure}
 \hfill
\begin{subfigure}[t]{0.4\textwidth}
    \centering
    \caption{Flexibility aggregation with system loss compensation}\label{fig::performance::SLC} 
    \includegraphics[width=\textwidth]{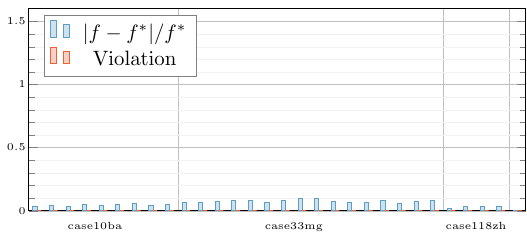}
\end{subfigure}
\caption{Performance of aggregating flexibility in a distribution system, considering one TSO connected with $29$ distribution systems, including $9$ case10ba, $16$ case33mg and $4$ case118zh}\label{fig::performance}\vspace{-12pt}
\end{figure} 

The results of the post-coordination calculation, summarized in \cref{fig::performance}, demonstrate significant improvements in coordination accuracy achieved by the loss compensation method compared to the LinDistFlow model. In \cref{fig::performance::SLC}, we can see that the compensation method effectively reduces cost $f$ discrepancies across all networks, aligning closely with centralized coordination cost $f^*$. In contrast, the LinDistFlow model exhibits larger cost deviations, as depicted in \cref{fig::performance::LDF}. In particular, the case33mg, which is more sensitive to power losses due to its smaller size and lower demand, shows large cost deviations. This sensitivity underscores the limitations of the LinDistFlow model in accurately reflecting network characteristics under such conditions.

Additionally, the state variables within the distribution networks are resolved during the post-coordination calculation. Using the compensation method ensures that these variables remain within physical constraints, leading to no violations in \cref{fig::performance::SLC}. In contrast, the results in \cref{fig::performance::LDF} highlight the violations that occur when using the LinDistFlow model. Voltage lower-bound violations are consistent with expected behavior when \glspl{der} attempt to meet more demand, as higher voltage magnitudes are required. Reactive power constraints emerge as the most frequently violated limits, underscoring the critical role of reactive power in maintaining efficient power transmission and stable voltage levels.
\section{Conclusion}\label{sec::conclusion}
The present paper analyzes the limitations of the LinDistFlow model for flexibility aggregation, highlighting its inherent approximation errors in both flexibility aggregation and \gls{tso}-\gls{dso} coordination. To address these issues, we proposes a loss compensation method that leverages the computational efficiency of the LinDistFlow model while enhancing its accuracy for aggregation and coordination tasks. Simulation results demonstrate that the proposed method significantly reduces infeasibility and aligns more closely with reference solutions, thereby improving the practicality of LinDistFlow-based aggregation.
\vspace{-5pt}




\bibliographystyle{IEEEtran}
\bibliography{root}
\end{document}

%% file: macros.tex
\usepackage{amssymb}
\usepackage{amsmath,amsthm,amsfonts}\usepackage[english]{babel}
\usepackage{dsfont}

\usepackage{xspace}
\usepackage{cases}
\usepackage[dvipsnames]{xcolor}
\usepackage{graphicx}
\usepackage[font=footnotesize]{caption}
\usepackage[font=footnotesize]{subcaption}
\usepackage{epstopdf}
\usepackage{upgreek}
\usepackage{tikz}
\usetikzlibrary{shapes.geometric, arrows}
\usepackage{booktabs}
\usepackage{array}
\usepackage{blindtext}
\usepackage{multirow}
\usepackage{bbold}
\usepackage{soul}
\usepackage{enumerate}
\usepackage{blindtext}
\usepackage{siunitx}
\usepackage{pgf}
\usepackage{lipsum}
\usepackage{hyperref}
\usepackage[acronym,hyperfirst=true]{glossaries}
\usepackage{enumitem}
\usepackage{setspace}
\hypersetup{
  colorlinks=true,
  citecolor=Violet,
  linkcolor=Red,
  urlcolor=Blue}
\setitemize[1]{itemsep=4.5pt,partopsep=1pt,parsep=\parskip,topsep=4.5pt}
\usepackage{algorithm,algorithmic}

\setkeys{glslink}{hyper=false}

\newtheorem{ass}{Assumption}

\newtheorem{propo}{Proposition}

\newtheorem{rema}{Remark}

\newacronym{opf}{OPF}{optimal power flow}
\newacronym{qp}{\textsc{qp}}{quadratic program}
\newacronym{nlp}{\textsc{nlp}}{nonlinear programming}
\newacronym{rapidpf}{rapid\textsc{pf}}{rapid prototyping for distributed Power Flow}
\newacronym{admm}{\textsc{admm}}{Alternating Direction Method of Multipliers}
\newacronym{aladin}{\textsc{aladin}}{Augmented Lagrangian based Alternating Direction Inexact Newton method}
\newacronym{ocd}{\textsc{ocd}}{Optimality Condition Decomposition}
\newacronym{app}{\textsc{app}}{Auxiliary Problem Principle}
\newacronym{sqp}{\textsc{sqp}}{Sequential Quadratic Programming}
\usepackage{titlesec}
\titlespacing*{\section}{0pt}{*0.8}{*0.8}
\titlespacing{\subsection}{0pt}{*0.8}{*0.8}

\definecolor{RB}{rgb}{.1,.4,.9}
\definecolor{VH}{rgb}{0.,.8,.4}
\definecolor{revised}{rgb}{.2,.8,.1}

\newacronym{tso}{TSO}{transmission system operator}
\newacronym{dso}{DSO}{distribution system operator}
\newacronym{der}{DER}{distributed energy resource}
\newacronym{pcc}{PCC}{point of common coupling}
\newacronym{bim}{BIM}{bus injection model}
\newacronym{ess}{ESS}{energy storage systems}
\newacronym{soc}{SOC}{state of charge}
\newacronym{itd}{ITD}{integrated transmission-distribution}

\makeglossaries
\usepackage{cleveref}
\crefname{figure}{Fig.}{Figs.} 
\Crefname{figure}{Fig.}{Figs.} 
\crefname{equation}{Eq.}{Eqs.} 
\Crefname{equation}{Eq.}{Eqs.} 
\usepackage{subcaption}
\usepackage{makecell}

\renewcommand{\labelitemi}{--}

\newcommand{\abs}[1]{\left|#1\right|}
\newcommand{\parens}[1]{\left(#1\right)}

